# Formation and Thermodynamic Behavior of THF-Water Hydrates in Confined Mesoporous Media

Armin Mozhdehei[1,*]

Oriana Osta[2]

Thomas Marescot[2]

Arnaud Desmedt[2,3]

Denis Morineau[1]

Christiane Alba-Simionesco[2]

1. Institute of Physics of Rennes, CNRS-University of Rennes 1, UMR 6251, F-35042 Rennes, France.
2. Laboratoire Léon Brillouin, Gif-sur-Yvette, France.
3. Institut des Sciences moléculaires, UMR5255 CNRS – Université de Bordeaux, F-33405 Talence Cedex, France

* armin.mozhdehei@univ-rennes.fr

## Abstract

Tetrahydrofuran (THF) is a benchmark guest for probing clathrate hydrate thermodynamics because a stoichiometric aqueous solution ($THF.17H_2O$) forms structure-II (sII) hydrate at ambient pressure with a well-defined dissociation temperature. Here, we combine differential scanning calorimetry (DSC) and wide-angle X-ray scattering (WAXS) in bulk and confined media to resolve how composition, pore filling, and cooling rate govern hydrate formation in SBA-15 mesoporous silica. Bulk DSC establishes mass-balanced enthalpies for ice and sII hydrate and confirms reversible dissociation/melting temperatures. In confinement, the heating traces separate into a Gibbs-Thomson depressed ice melt (= −14.7 ± 0.2 °C), an in-pore hydrate dissociation (= −13.2 ± 0.2 °C). Confined hydrate appears only when two criteria are met: near-percolating filling (φ = 1.0 – 1.1 cm3/g) and sufficient THF (≥ 1:16 mol:mol). Cooling-rate experiments (1.0 vs 0.5 °C/min) demonstrate that slower precooling increases the confined-hydrate fraction and reduces confined ice without shifting equilibrium temperatures: at φ = 1.1, the hydrate enthalpy rises by ~60% at 1:11 and ~54% at 1:14, but by ≤ 17% at 1:16. Temperature-cycling tests show invariant reheating peak positions, indicating that capillarity and composition, rather than kinetic history, fix the liquidus and dissociation temperatures. WAXS indicates that the phase formed in pores is crystallographically identical to bulk sII. Finally, the variation of melting points ($\Delta T_m$) plotted against inverse pore radius follows the Gibbs-Thomson law for both ice melting and hydrate dissociation, quantitatively linking the observed shifts to crystalline size and clarifying how confinement, cooling rate, and composition govern the competition between hydrate formation and water crystallization.

# 1. Introduction

Clathrate hydrates are crystalline inclusion compounds in which a hydrogen-bonded water lattice hosts small guest molecules. Natural gas hydrates, which contain small alkanes like methane and propane, have been extensively studied due to their importance in energy-related applications and natural sedimentary deposits.[1, 2] In porous media, predicting hydrate equilibrium and decomposition conditions is a key challenge in understanding hydrate occurrence, stability, and recovery.[1, 3] In this context, laboratory experiments have been conducted on methane hydrates in porous media, such as sandstone, clays, and glass beads, to replicate natural conditions.[3] Observations show that hydrate decomposition temperatures shift downward by up to 1.5 K, which is attributed to the varying confining pore sizes, ranging from one to a few tens of micrometers. Stronger confinement effects have been achieved using mesoporous synthetic silica gel with a mean pore diameter of 14 nm as the confining matrix.[4] In such confined systems, the pressure for both hydrate-ice-gas and hydrate-liquid water-gas equilibria was systematically higher than in bulk systems. This confinement-induced shift in the pressure-temperature equilibrium line can be explained using classical thermodynamic models, where the primary influence comes from the capillary pressure effect, which is a function of the mean pore size, surface energy, and wetting angle.[5] Consistently, calorimetric studies of nanoconfined methane gas hydrates have shown that the hydrate dissociation temperature is lowered by approximately 10–30 K compared to the bulk state.[4] Furthermore, these measurements have revealed unusual thermal behaviors: the initial dissociation of pore hydrates is restricted because the system is encapsulated by a stabilizing ice layer. As a result, most of the hydrate (~70%) ultimately dissociates at a higher temperature than thermodynamically predicted, only when the ice melts. This phenomenon has practical implications, such as in gas fields or pipelines, but it also significantly complicates fundamental studies of confinement effects.

For this reason, polar organic molecules, which also promote clathrate hydrate formation, provide a valuable alternative as model systems. Among them, Tetrahydrofuran (abbreviated as THF hereafter) is one of the most extensively investigated guest molecules and serves as a widely used clathrate hydrate model system.[6, 7] Unlike gas clathrate hydrates, which require high pressure, THF hydrates can form under relatively mild experimental conditions, making them particularly suitable for fundamental clathrate hydrate studies. The three common clathrate crystalline forms are cubic structures I, II, and H. THF molecules stabilize the structure of the sII framework by occupying its large cages,[1, 8] proven by X-ray[9] and neutron diffraction.[10, 11] In the structure-II (sII) clathrate, each unit cell comprises 136 water molecules forming 24 cages: 16 small pentagonal dodecahedra ($5^{12}$) and 8 larger 16-hedra ($5^{12}6^{4}$).[2, 12, 13] Also, sII hydrates have a cubic lattice parameter (with $a$ ~ 17.3 Å), and the average van der Waals radii of cages are *ca*. 3.7 Å for $5^{12}$ and 4.2 Å for $5^{12}6^{4}$.[2, 8, 9, 14] THF is particularly advantageous due to its ability to form a sII hydrate from a stoichiometric aqueous solution (THF.17$H_2O$) at ambient pressure, exhibiting a well-defined melting/dissociation temperature. This behavior produces a univariant hydrate-solution equilibrium boundary in the temperature-pressure (T-P) phase diagram, as the fixed guest-to-water ratio in the clathrate constrains the equilibrium composition.[15, 16]

At lower temperatures, the hydrate-solution line terminates at a well-defined four-phase point (hydrate–ice Ih–solution–gas) at 272.06 ± 0.02 K (≈1.1 kPa), which anchors the phase diagram and provides a practical benchmark for validating calorimetric protocols such as controlled differential scanning calorimetry (abbreviated as DSC hereafter) interrogation of hydrate thermodynamics without high-pressure hardware.[7, 16-19] From a broader perspective, THF is a thermodynamic promoter: by stabilizing sII, it lowers formation pressures and raises formation temperatures for mixed hydrates with gases such as $H_2$, $CO_2$, or $CH_4$, motivating its use in storage/separation concepts and in fundamental datasets for model development.[6, 20-22]

Rigorous descriptions of the THF-water phase behavior now couple liquid-phase activity models (e.g., NRTL) with a cubic equation of state (Peng–Robinson) for vapor fugacities and the van der Waals-Platteeuw framework with Kihara potentials for cage occupancy, yielding a thermodynamically self-consistent baseline for interpreting DSC across vapor-liquid, hydrate-liquid, and ice-liquid equilibria.[21] These models, together with spectroscopic and simulation evidence, indicate that host-guest stabilization in THF hydrates is not purely dispersive: in addition to van der Waals forces, specific guest-host interactions help maintain the water framework.[23, 24] Yet formation remains a kinetic, molecular-scale problem. Bulk-sensitive x-ray Raman spectra measurements on THF-water are consistent with a stochastic formation picture ("local structuring"), in which hydrate formation is interpreted as occurring when transient local configurations become compatible with the clathrate structure, without evidence for persistent, spectroscopically resolvable precursors.[25] At the same time, alternative views such as cluster-nucleation allow for very short-lived precursor clusters in supercooled mixtures.[2, 25] Molecular simulations further show that the THF ether oxygen can transiently hydrogen-bond to cage waters, generating Bjerrum defects, thermally activated violations of the ice rules with either two hydrogens (D-defect) or none (L-defect) between adjacent oxygens, which subtly perturb mechanical and transport properties near dissociation.[26, 27] This disorder in the hydrogen-bond network of the water framework is well established, as shown by dielectric spectroscopy,[28] proton ($^1$H) NMR,[29] solid-state deuterium ($^2$H) NMR studies,[30-32] and quasielastic neutron scattering (QENS).[33]

While molecular modeling and spectroscopy delimit the microscopic pathways to the formation of clathrate, calorimetry offers a clean way to separate kinetics from thermodynamics in the THF-water mixture. Calorimetrically, THF hydrate offers clean standards: at atmospheric pressure, its reversible melting/dissociation temperature near $277 \pm 0.2$ K,[16, 17, 19] the enthalpy and entropy of fusion,[16-18, 34] and the comparative heat capacities of hydrate versus components have been quantified, enabling rigorous energy/entropy balances and providing reference values for validating DSC protocols and interpreting metastability.[18]

While the stability of natural gas hydrate confined in porous media has been investigated,[3-5, 35] the behavior of THF clathrates in such environments remains to be documented. This is particularly relevant because THF allows us to explore a distinct physical scenario. Indeed, at ambient pressure, ice melts at a lower temperature than the THF clathrate.[16, 18] In calorimetric measurements, two endothermic peaks are typically observed upon heating, corresponding to the successive melting of ice and THF hydrate. Depending on the composition and thermal history, the ratio of ice to hydrate formed can vary.[16, 18] This differs from gas hydrates, where the first thermal event is usually associated with clathrate dissociation into ice and gas, and the second with ice melting.[4] This means that the anomalous stabilization observed for some gas hydrates,[4] due to ice encapsulation does not arise in the same way in the THF–water system. The presence of two independent melting processes in THF is particularly advantageous for fundamental studies, as it allows for the examination of confinement effects on phase behavior, as well as the competition between ice and THF clathrate formation.

Taking advantage of the prototypical nature of THF hydrate to deepen the fundamental understanding of clathrate formation, Xue et al. recently determined the critical nucleus size by observing the effects of introducing nanoparticles of various controlled sizes.[36] They established a correlation between the critical nucleus radius $r_c$ and the inverse degree of supercooling $\Delta T$, expressed as $r_c = K/\Delta T$ with $K$ = 300 °C nm. Under a conventional cooling rate condition (1 °C/min), bulk THF hydrate forms at a supercooling of $\Delta T \approx 26$°C. Meanwhile, Zhang et al. demonstrated that homogeneous crystallization occurs at a higher supercooling of $\Delta T \approx 40$°C.[16] Applying the aforementioned relationship, the critical nucleus radius can be estimated at approximately 12 nm and 7.5 nm, respectively, i.e., sizes typical of nanoporous materials. Therefore, these recent findings by Xue et al. underscore the need for a detailed investigation of the effect of nanoscale confinement on THF hydrate formation and stability. Such a study was initiated in the pioneering work by Zakrzewski and Handa[37] for THF hydrate confined in Vycor with a 4.2 nm pore radius. This work first demonstrated the possible formation of THF hydrate in nanoporous geometries and showed that confinement affects the melting of THF hydrate to a similar extent as it does for ice, with a reduction in the melting temperature of about 8.5 K. While interesting, this study lacks fine control over the pore geometry and raises open questions, such as the impact of systematically varying the pore size, composition, and pore filling. This study serves as the motivation for the current work.

In view of the foregoing, we used DSC on THF-water both in bulk and under nanoconfinement in highly regular mesostructured porous silicas (SBA-15). These porous matrices feature carefully designed cylindrical nanochannels, which are highly monodisperse in size and aligned in parallel. They therefore provide a unique experimental platform for investigating nanoscale confinement effects on phase behavior, while allowing fine and systematic tuning of pore size. As such, they have opened up important research avenues in areas such as the phase behavior of confined ice,[38,] [39] aqueous solutions,[40, 41] glass-forming liquids,[42, 43] binary organic mixtures,[44] and deep eutectic solvents.[45] In the present work, we sought to address five tightly linked questions: (i) how precise compositional control around the stoichiometric ratio governs phase selection; (ii) how the presence of excess liquid at high filling alters hydrate formation inside the pore network; (iii) how thermal history, including cooling rate and cycling, modulates the fraction of phases formed without shifting equilibrium temperatures; (iv) how the measured transition temperatures map quantitatively onto capillarity across pore sizes; and (v) what size limit constrains THF-water hydrate formation under confinement. By embedding each experiment within the established bulk phase boundaries and molecular pictures of guest–host interaction, we deliver a coherent, thermodynamically grounded reading of DSC thermograms that clearly separates equilibrium markers (invariant reheating temperatures) from signatures of kinetic effects (rate-dependent exotherms and peak areas).

# 2. Materials and methods

## 2.1. Samples

Tetrahydrofuran (THF; anhydrous, contains 250 ppm BHT as inhibitor, ≥99.9% purity) was obtained from Sigma-Aldrich (Merck) and used as received. Ultrapure deionized water (18.2 MΩ·cm at 21.8 °C) was produced on an ELGA PURELAB Classic DI system. SBA-15 mesoporous silica was synthesized by a surfactant-templated route analogous to prior works,[42, 43, 45-47] using the nonionic triblock copolymer Pluronic P123, $(EO)_{20}(PO)_{70}(EO)_{20}$, to form a hexagonally ordered array of aligned cylindrical channels. Pore metrics (i.e., pore size and pore volume) were obtained from $N_2$ adsorption isotherms: KJS analysis yielded an average mesopore diameter ~ 8.1 nm ($R_P \approx$ 4 nm), while HK analysis indicated a secondary microporosity centered ~ 1 nm; the pore volume was $\rho \approx 0.81 - 0.94\ cm^3/g$. To complete the pore-size series, we additionally purchased three SBA-15 materials (nominal mesopore diameters 4–10 nm) from Sigma-Aldrich, and verified their hexagonal mesostructural order by small-angle X-ray scattering (SAXS) through the presence of the (100) peak. All corresponding data and figures are provided in the Supplementary Information (Figs. S2a & S2b).

Prior to use, calcined powders of SBA-15 were dried for 18 h at 120 °C under primary vacuum. For DSC, between 7 and 9 mg of the dry matrix were packed in Tzero® hermetic aluminum pans and filled by liquid imbibition with either pure water or THF-water solutions delivered by syringe. The loaded volume was set between 0.7 and 1.1 $cm^3/g$ liquid volume per mass of matrix.

## 2.2. Differential Scanning Calorimetry (DSC)

Differential scanning calorimetry (DSC) was used to track hydrate/ice formation and dissociation/melting. Measurements were performed on a TA Instruments Q20 with $LN_2$ cooling, as in our team's prior works,[45, 48, 49] with temperature and heat-flux calibrated using indium. Samples were sealed in aluminum hermetic pans with a matched empty reference, and all runs were conducted under a dry He purge. Each run held the sample at 20 °C for 1 min, cooled to -40 °C at 1.0 $°C.min^{-1}$ (or 0.5 $°C.min^{-1}$ where specified), held at -40 °C for 1 min, then heated back to 20 °C at 2.0 $°C.min^{-1}$. Crystallization exotherms of THF clathrate hydrate and ice were recorded on cooling, and the corresponding dissociation/melting endotherms on heating. Pan sealing and reproducibility were confirmed by at least two identical cooling-heating loops yielding indistinguishable peak temperatures and shapes, and cell integrity was checked by weighing samples immediately before and after each experiment (see Supplementary Table S1 for mass differences). The slower cooling rate (0.5 $°C.min^{-1}$) was used for investigating the effect of cooling rate on the formation of the hydrate. Confined (in-pore) versus external (out-pore) phase transitions were distinguished by their characteristic temperatures and peak profiles. Heat-flow–temperature thermograms were processed in TA Instruments Universal Analysis, in which onset temperatures were determined by the tangent-intersection method, and peak temperatures were taken at the heat-flow maxima. Also, the transition enthalpies were obtained by baseline-subtracted integration and normalized to sample mass to estimate relative phase fractions.

# 3. Results and Discussion

## 3.1. Phase behaviour in bulk systems

Fig. 1 shows the DSC thermogram of THF-water mixtures spanning 1:11, 1:14, 1:16, and 1:17 (THF:$H_2O$, mol:mol) in bulk condition yielded two reproducible endotherms on heating: a lower-temperature peak attributable to melting of ice (around 0 °C ± 0.25) and a higher-temperature peak (around 5 °C ± 0.25) due to melting of the THF clathrate hydrate (structure II, THF.17$H_2O$), while only one exothermic peak (around -18 °C ± 0.25) is observed upon cooling, as established before.[16, 18]

In the DSC thermograms, heat flow is displayed as specific power (W/g). Therefore, integrating a peak gives an energy per unit mass (J/g). To report the total enthalpy (ΔH) associated with each transition in a given pan ($\Delta Q$, in J), we multiplied the integrated value (J/g) by the initial sample (total) mass loaded in that pan. Using this procedure, the melting enthalpies for ice across the four mixtures (1:11 → 1:17) are 0, 0.13, 0.36, and 0.38 J, while those for the hydrate are 1.10, 1.01, 0.88, and 0.87 J. As expected, the peak areas increase with the amount of the corresponding solid phase present. To convert each peak area $\Delta Q$ (i.e., enthalpy) to the mass of the solid that melted at a specific temperature, we divided by the appropriate enthalpy of melting. For ice, it equals $\Delta H_{\mathrm{fus,ice}}$= 333.55 J/g, and is accurately known.[50, 51] For the clathrate, we used a per-mole value determined in the previous papers that is around $\Delta H_{\mathrm{fus,hyd}} \approx$ 99.5 kJ/mol[16-18, 34] and the hydrate formula mass at its stoichiometric composition (THF.17$H_2O$) equals $M_{\mathrm{hyd}}$ = 378.37 g/mol, giving a specific latent heat $\Delta H_{\mathrm{fus,hyd}}$ = 261.91 J/g.

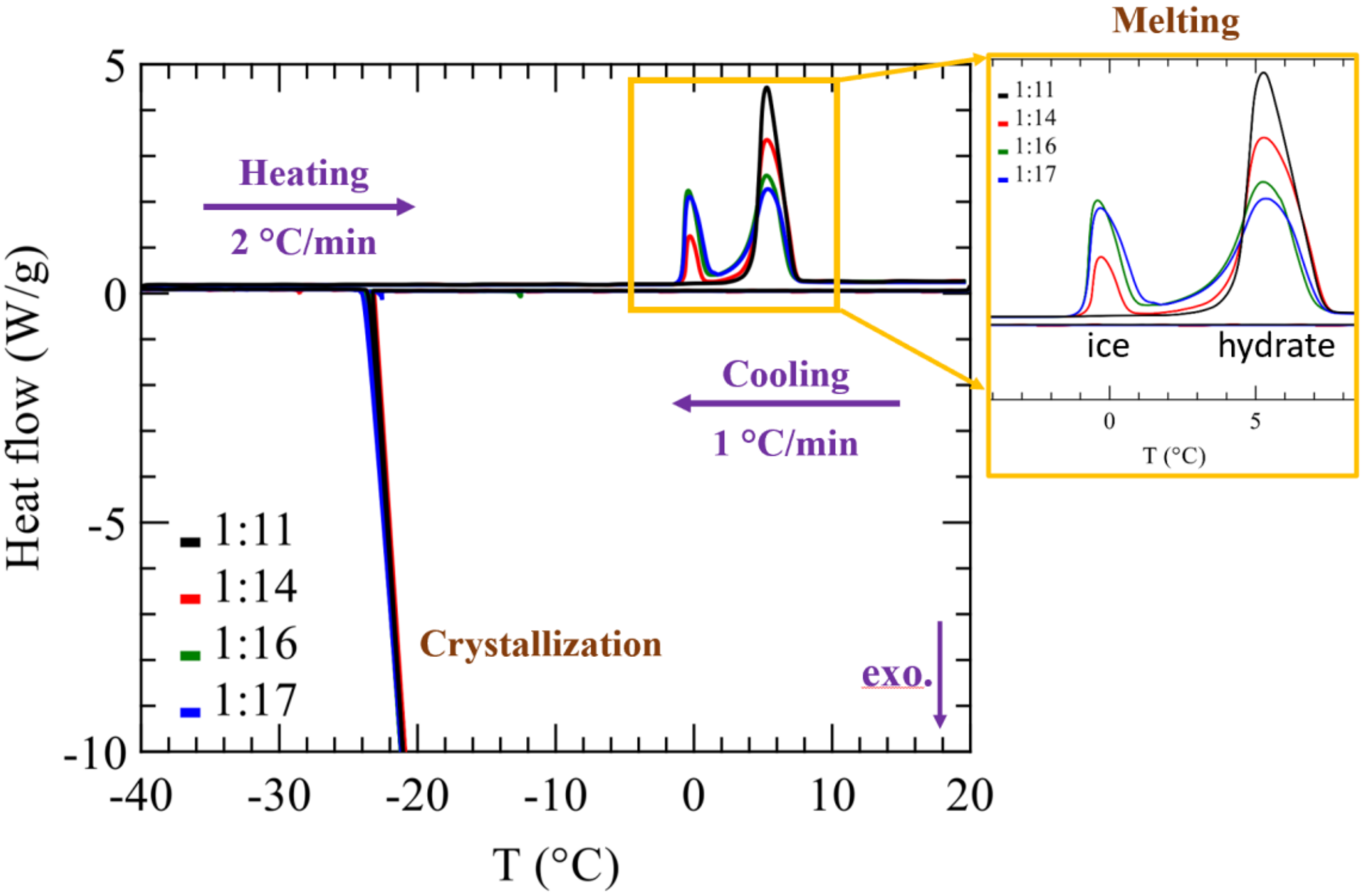

*Fig. 1. DSC thermogram of bulk THF-water at four compositions (molar ratios 1:11, 1:14, 1:16, 1:17). Samples were cooled at 1 °C.min$^{-1}$ and reheated at 2 °C.min$^{-1}$ (exothermic down). The inset magnifies the heating endotherms: the left peak corresponds to ice melting (near 0 °C) and the right peak to sII THF-hydrate dissociation (near 5 °C). Peak areas vary systematically with composition; hydrate dominates at higher THF content (1:11), while ice increases as mixtures are diluted (1:17).*

To quantify the proportions of each phase formed, we used a method adapted from the approach described in [Ref. 15]. [18] This reference demonstrated the reversible melting of the THF hydrate near 277 K and reported an equilibrium melting enthalpy of approximately 99.5 kJ mol$^{-1}$. The hydrate's stoichiometry results in a composition of 19.06 wt% THF and 80.94 wt% $H_2O$. Applying these constants to our peak integrals yields the phase amounts present in the following table. It offers a direct measurement of the amounts of THF and water that were actually converted into solid phases (i.e., ice and hydrate), compared to the masses initially available in the solutions. These values are mass-balanced against what each pan contained a priori (from the homogeneous starting compositions).

*Table 1. DSC-derived partitioning in THF-$H_2O$ mixtures: heating integrals ($\Delta Q_{ice}$, $\Delta Q_{hydrate}$) convert to masses of ice and sII THF.17$H_2O$ hydrate; rightmost columns report fractions of initial THF and $H_2O$ converted into solid phases (i.e., ice and hydrate).*

| Mixture (THF:$H_2O$) | $\Delta Q_{ice}$ (J) | $\Delta Q_{hydrate}$ (J) | Water as ice (mg) | Total hydrate (mg) | THF in hydrate (mg) | Water in hydrate (mg) | THF conv / THF avail | Water conv / Water avail |
|---|---|---|---|---|---|---|---|---|
| 1:11 | 0.00 | 1.10 | 0.0 | 4.20 | 0.87 | 3.40 | 0.70 | 1.00 |
| 1:14 | 0.13 | 1.01 | 0.39 | 3.85 | 0.73 | 3.12 | 0.69 | 0.94 |
| 1:16 | 0.36 | 0.88 | 1.06 | 3.36 | 0.64 | 2.72 | 0.66 | 0.98 |
| 1:17 | 0.38 | 0.87 | 1.25 | 3.32 | 0.63 | 2.69 | 0.65 | 0.95 |

A clear composition-dependent redistribution emerges. As THF availability increases, a larger fraction of THF is trapped in clathrate at the expense of ice. The THF mass sequestered in hydrate evolves 0.87, 0.73, 0.64, and 0.63 mg for 1:11, 1:14, 1:16, and 1:17, respectively, corresponding to ≈ 70%, 69%, 66%, and 65% of the initial amount of THF. The associated hydrate-bound water is 3.40, 3.12, 2.72, and 2.69 mg across the same series, with the remaining water appearing as bulk ice at the melting onset and a small residual remaining in solution or disappearing by evaporation. For the 1:14 mixture, initially loaded with 3.73 mg $H_2O$ and 1.07 mg THF, the DSC-derived phase partitioning assigns ~0.39 mg of $H_2O$ to ice and ~3.12 mg to hydrate. Thus, ~3.51 mg of water participates in solid-phase formation, leaving ~0.22 mg either in the residual liquid phase or lost through evaporation. THF in hydrate is ~ 0.73 mg, leaving ~ 0.33 mg in the liquid, both within the initial inventories.

In fact, at the most water-deficient composition (1:11), the THF conversion is maximal: the heating trace shows no ice-melting endotherm, implying that essentially all $H_2O$ is incorporated into sII clathrate at the dissociation onset. Yet the THF mass recalculated from the hydrate enthalpy is ~30% lower than the initial amount. Two loss pathways were considered. (i) Loss through the DSC pan is unlikely: pre- and post-run weightings of the crimped crucibles differed by only ~ 0.05 mg, far below the inferred THF deficit, ruling out in-run leakage as the dominant source. (ii) Pre-sealing handling/evaporation during transfer by syringe is plausible given THF's volatility; a fraction of the guest can be lost before the pan is hermetically closed.

We also tested whether "missing" THF might simply remain as neat liquid THF in the crucible. The program was extended to −120 °C to search for the pure THF melting endotherm (expected near −108.4 °C)[52, 53] for the 1:14 concentration. As shown in Fig. 2, no such event was detected, excluding a persistent neat THF phase. Taken together, the mass closure, the absence of an ice-melting peak, and the lack of a pure THF melting signal indicate that the apparent ~30% THF shortfall in the 1:11 case arises upstream of sealing (minor evaporative loss during loading due to the higher volatility of THF), not from leakage during the DSC scan nor from unincorporated liquid THF in the cell.

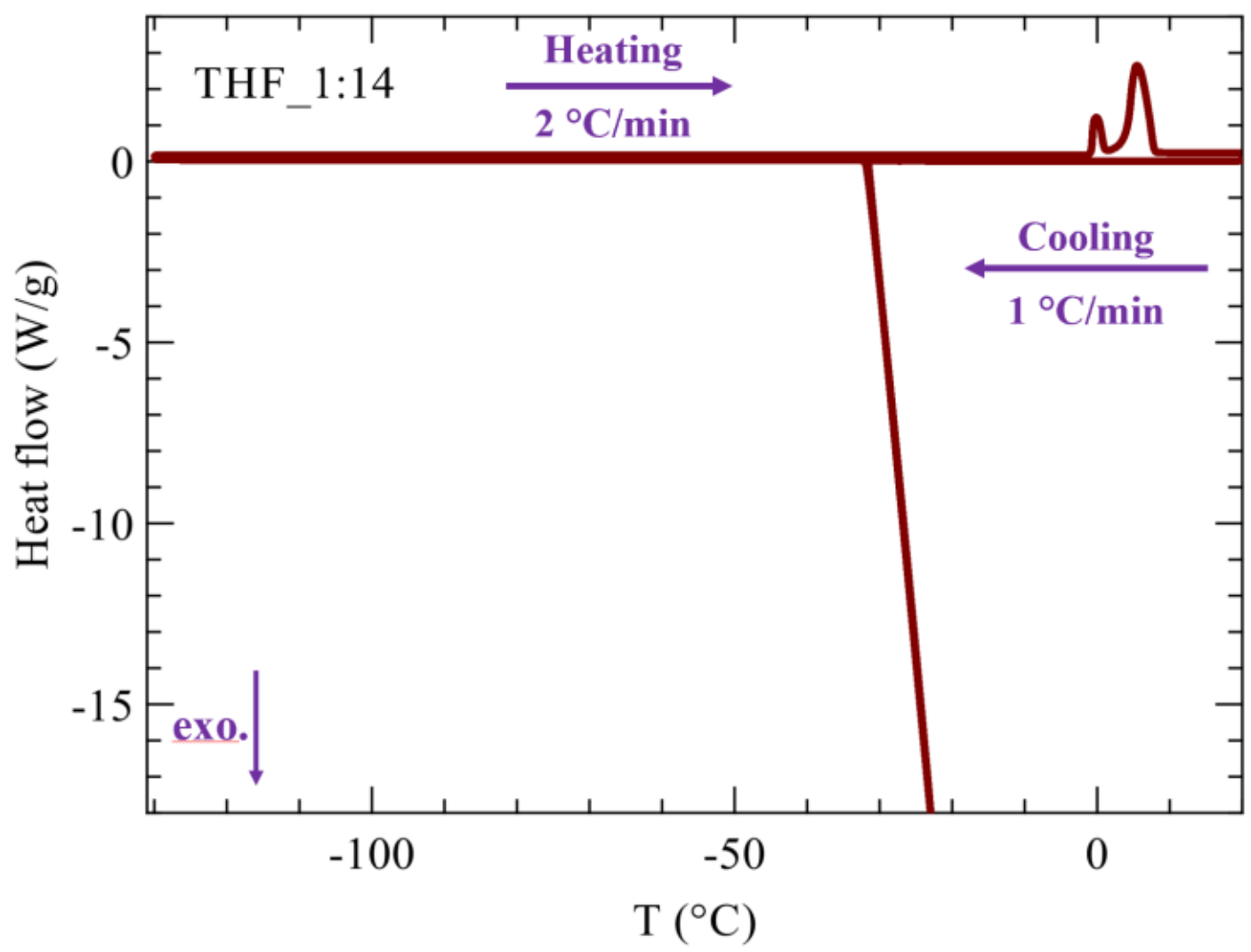

*Fig. 2. DSC of bulk THF-water (1:14). Cooling: 1 °C.min$^{-1}$ to −130 °C; heating: 2 °C.min$^{-1}$ (exothermic down). No endotherm appears near −108.4 °C, ruling out residual neat THF. Inset: heating endotherms, ice melting (~ 0 °C) and sII THF–hydrate dissociation (~ 5 °C).*

## 3.2. Phase behaviour in confined systems

These bulk insights set the stage for interrogating how thermal itinerary and confinement modulate what DSC records. In this way, the phase transition of the THF-water mixtures in confined media was evaluated using SBA-15. Fig. 3 shows the thermograms of the THF:water at four tested mole fractions and at different filling fractions in the pore with a diameter of 8 nm. Across all compositions (1:11 → 1:17), the filling fraction φ (cm$^3$/g) defines whether the system contains only confined (i.e., in-pore) solution or both confined and excess phases (i.e., out-pore). This distinction clarifies where phase transition occurs and, consequently, which thermal signatures are detected. In Fig. S3 (SI), we use pure water in SBA-15 with d = 8 nm as a calibration to determine pore volume and to establish the confined-ice reference under identical matrix, filling fraction, and scan conditions. We observed that for $\varphi < 0.9$ the pores contain only confined liquid (no excess outside the mesochannels), and the reheating trace shows a single dominant endotherm at −14.7 ± 0.2 °C, the expected classical Gibbs–Thomson depressed melting of confined ice in SBA-15.[38] This control unambiguously fixes the confined-ice melting temperature for our system; consequently, any endotherm at this temperature in THF-water runs is assigned to confined ice, not sII THF hydrate. For cooling, we observed one exotherm at a bit higher temperature than the bulk, indicating reduced supercooling under confinement. There is no systematic change based on the initial concentration for the two exothermic peaks assigned to the bulk (i.e., out-pore) and confined (i.e., in-pore) liquid crystallization on cooling, as expected for heterogeneous nucleation governed by stochastic nucleation kinetics rather than precursor-based scenarios, as it was shown by X-ray Raman measurements on THF–water mixtures.[25]

When φ approaches 1.0 cm$^3$/g, the SBA-15 becomes overfilled, producing excess liquid. This results in a distinct, higher-temperature dissociation peak, observed around -13.2 ± 0.2 °C in the heating curve for THF-richer mixtures (1:11–1:16), signalling hydrate formation within the pores. A second peak at 3.8 ± 0.2 °C corresponds to a hydrate formed outside the pores. With a higher overfilling ($\varphi \approx 1.1$), a hydrate dissociation peak becomes more intense, which is consistent with continuous nucleation sites throughout the pore network. This can also be rationalized by chemical activity at the pore entrances. When excess liquid is present outside the pores, the external reservoir pins the water/THF activities at the pore entrances throughout the process. This buffers the in-pore composition as phases forms, sustaining conditions favorable to hydrate growth along the mesochannels. At $\varphi \leq 0.9$, by contrast, the in-pore liquid behaves as a closed inventory: once formation begins, local composition drifts and phase selection becomes more sensitive to stochastic pathways rather than to a fixed activity boundary. A further contribution may come from preferential THF partitioning near the silica surface. If part of the THF is associated with interfacial silanol-rich regions, the effective THF concentration in the pore-center liquid may be reduced. This effect should be strongest at low filling, where the surface-to-volume ratio is highest, and could shift the pore-core composition away from hydrate stoichiometry, thereby suppressing in-pore hydrate formation. A consistent picture emerges in the study of cyclopentane-loaded activated carbons. Mallek et al. showed that hydrates form predominantly from guests located in large pores or on particle surfaces, and that providing excess (oversaturated) guests dramatically increases the converted fraction, whereas guests trapped in small pores hardly participate.[54] Regarding the second higher-temperature peak, the bulk-like peak appears to be approaching the $T_m$ of bulk THF while in confinement the endothermic peak is also now clearly separating (Gibbs–Thomson depressed) from ice.

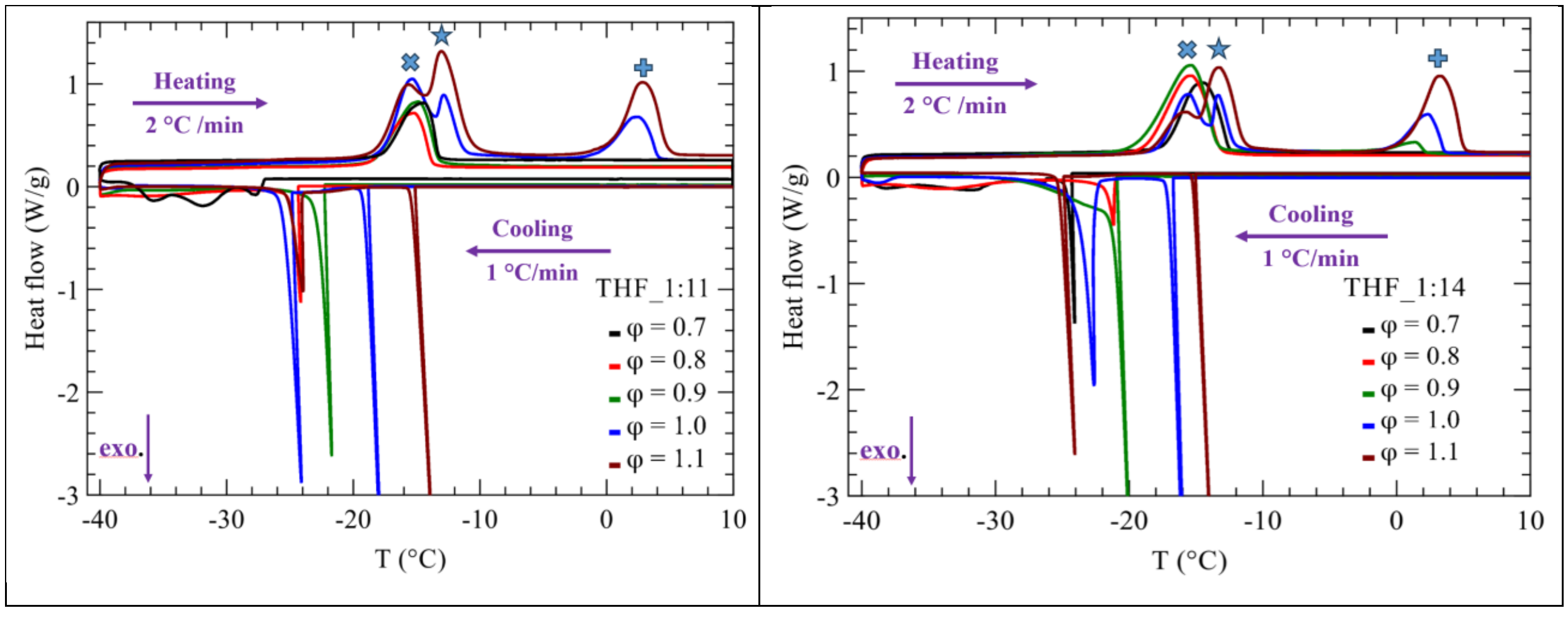

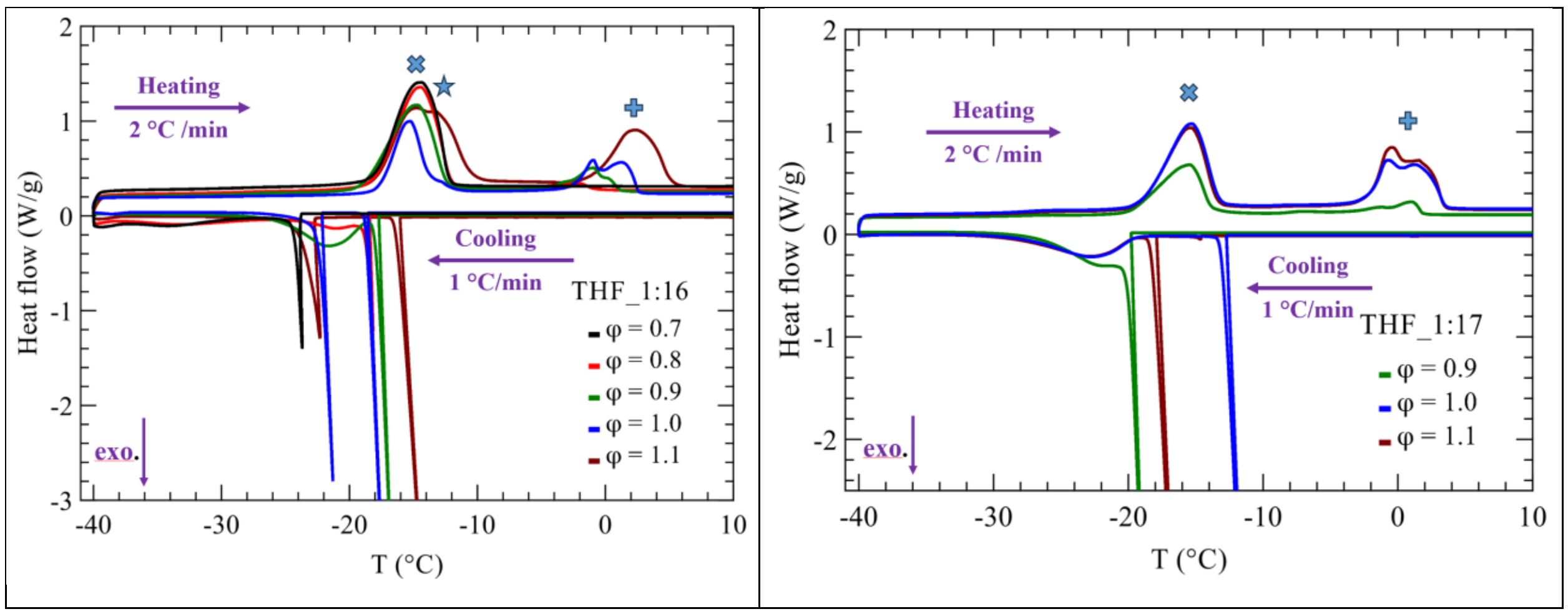

*Fig. 3. DSC thermograms of THF–water mixtures confined in SBA-15 at four compositions, THF:$H_2O$ = 1:11, 1:14, 1:16, and 1:17, and at different filling fractions φ. Samples were cooled at 1 °C $min^{-1}$ and reheated at 2 °C $min^{-1}$ (exothermic downward). On heating, the symbols identify the different phase transitions: × indicates the melting of **confined ice**, ★ the dissociation of **confined hydrate**, and + the dissociation of the **bulk/external hydrate phase**. Confined ice melts near −14.7 °C, confined hydrate dissociates near −13.2 °C, and the bulk-like hydrate dissociates near +4 °C. The figure highlights the emergence of the confined-hydrate signal only at sufficiently high filling fractions and THF contents, while the 1:17 composition remains dominated by confined ice.*

At the most dilute nominal composition (1:17), there is a dissociation peak outside the pore along with the ice melting peak; however, no hydrate dissociation is observed at any filling fraction in the porous medium, including φ ≥ 1.0; the thermograms are therefore dominated by confined ice freezing/melting. While sII stoichiometry (THF.17$H_2O$) would make 1:17 the only strictly stoichiometric mixture for hydrate formation, a small but reproducible THF loss in our sample preparation, consistent with evaporation/partitioning during loading and cycling, shifts the effective composition toward water-rich conditions. As a result, the 1:17 samples fall outside the hydrate-forming window and serve as an ice-dominated reference. In the water-deficient series (1:11–1:16), where the post-loss composition lies closer to the hydrate domain, the filling-fraction dependence is clear and reproducible: hydrate signatures appear only when both criteria are satisfied: (i) adequate pore filling (φ ≈ 1.0 - 1.1) to establish continuous formation pathways and (ii) sufficient THF content (≥1:16). Below either threshold, the endotherm near the ice temperature is correctly assigned to confined ice rather than hydrate.

### 3.3. Effects of the cooling rate on the clathrate formation

Building on the filling-fraction results, the effect of the cooling rate was assessed and shown in Fig. 4 by varying from 1 °C/min to 0.5 °C/min while the heating rate was kept constant at 2 °C/min to avoid a bias change in the endothermic peak areas. Since the 1:17 concentration did not show hydrate formation within the porous medium, we proceeded with three other THF:$H_2O$ concentrations, namely 1:11, 1:14, and 1:16, at filling fractions of $\varphi \approx 1.0$ and $\varphi \approx 1.1$. With a constant heating rate, variations in the endotherm areas directly reflect cooling rate effects on the phase fractions created during cooling. Comparing the thermograms in Fig. 4, green and black for $\varphi = 1.0$, and red and blue for $\varphi = 1.1$, shows qualitatively that slower pre-cooling favors hydrate formation while suppressing confined ice. Consistently, on reheating, the endotherm attributed to hydrate dissociation in SBA-15 appears at −13.2 ± 0.2 °C and its integrated area increases markedly after cooling at 0.5 °C/min, whereas the ice-melting peak centered at −14.7 ± 0.2 °C recedes.

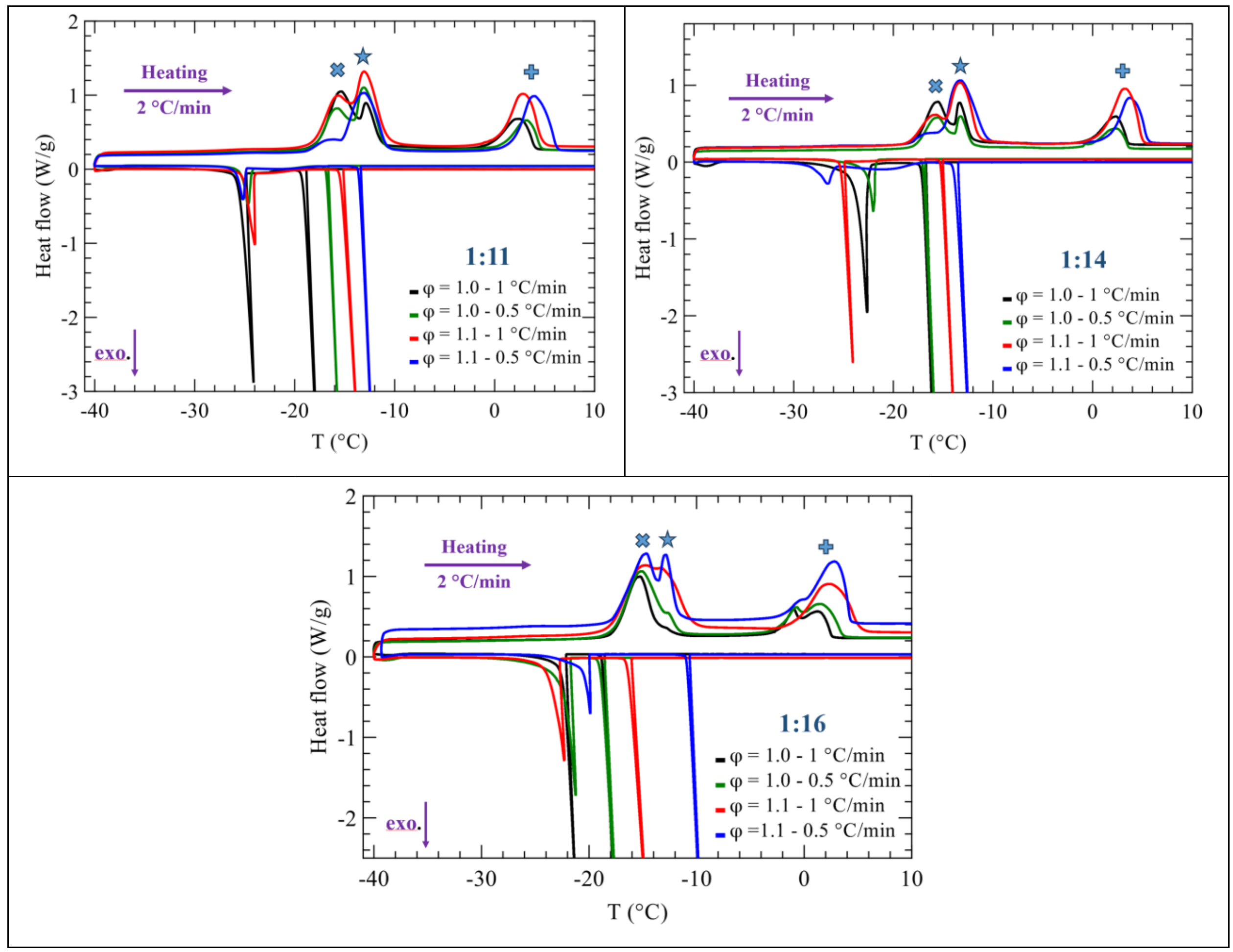

*Fig. 4. Cooling-rate effect for THF-water in SBA-15 at φ=1.0 and 1.1 for three compositions (1:11, 1:14, 1:16). Cooling: 1.0 or 0.5 °C.min$^{-1}$; heating fixed at 2 °C.min$^{-1}$ (exothermic down). Slower cooling increases the hydrate endotherm area while the confined-ice and confined-hydrate peak temperatures are essentially unchanged; the bulk-like hydrate peak shifts slightly to higher T. On heating, the symbols indicate the different phase transitions: × corresponds to the melting of confined ice, ★ to the dissociation of confined hydrate, and + to the dissociation of the bulk/external hydrate phase.*

As mentioned earlier, a second dissociation near 3.8 ± 0.2 °C is attributed to the hydrate formation outside the pores. This peak decreases slightly with the slower cooling rate by shifting toward higher temperature by around 0.6 °C, most strongly at 1:11 and 1:14 and still detectable at 1:16. By contrast, the exotherms on cooling exhibit the expected scatter for heterogeneous nucleation under confinement and show no systematic rate-dependent temperature shift, indicating that the principal role of the cooling rate is not to move the nucleation line but to alter the time available for phase selection and growth within the metastable window. Thermodynamically, hydrate is favored at these compositions, but kinetically it competes with rapid ice formation in mesopores. Reducing the cooling rate increases the time spent in the temperature interval favorable to hydrate formation during cooling. In the classical time–temperature–transformation (TTT) framework, this corresponds to a longer residence near the region where hydrate formation and growth are expected to be most efficient. Although the TTT curve was not determined experimentally here, this interpretation is consistent with the observed increase in hydrate formation at slower cooling. The additional time likely facilitates local mass redistribution and growth within the cylindrical pores, thereby reducing kinetic trapping into confined ice. Consistent with this picture, hydrate formation in SBA-15 is observed only when both criteria are satisfied, adequate filling (φ ≈ 1.0 – 1.1, to establish continuous pathways) and sufficient THF content (≥ 1:16). Within this thermodynamic window, reducing the cooling rate from 1.0 to 0.5 °C/min reproducibly shifts phase selection in favor of clathrate, evidenced by larger hydrate dissociation endotherms and weaker confined-ice signatures on reheating.

To quantify how the cooling pathway influences phase selection within SBA-15, we decomposed the two melting peaks using a fitting procedure and measured the integrated area of each peak (see Fig. 5a & 5b). Then, we determined the ratio of the confined hydrate enthalpy to the total endothermic enthalpy using equation 1.

$$f_{hyd} = \frac{\Delta H_{hyd}}{\Delta H_{hyd} + \Delta H_{ice}} \quad \textit{(equation 1)}$$

where $\Delta H_{hyd}$ and $\Delta H_{ice}$ are the enthalpies of confined-hydrate dissociation and confined-ice melting, respectively. The ratio $f_{hyd}$ is illustrated in Fig. 5c for two cooling rates (1.0 vs 0.5 °C/min), two filling fractions (φ = 1.0 and 1.1), and three compositions (1:11, 1:14, 1:16). The higher the $f_{hyd}$ value, the more clathrate formation is favored over ice. Fig. 5c visually confirms the two key findings from the previous section for a cooling rate of 1 °C/min: the formation of confined clathrate is promoted in slightly overfilled samples (φ = 1.1 outperforms φ = 1.0) and solutions with a higher initial THF concentration (1:11 over 1:16) further enhance clathrate formation over ice.

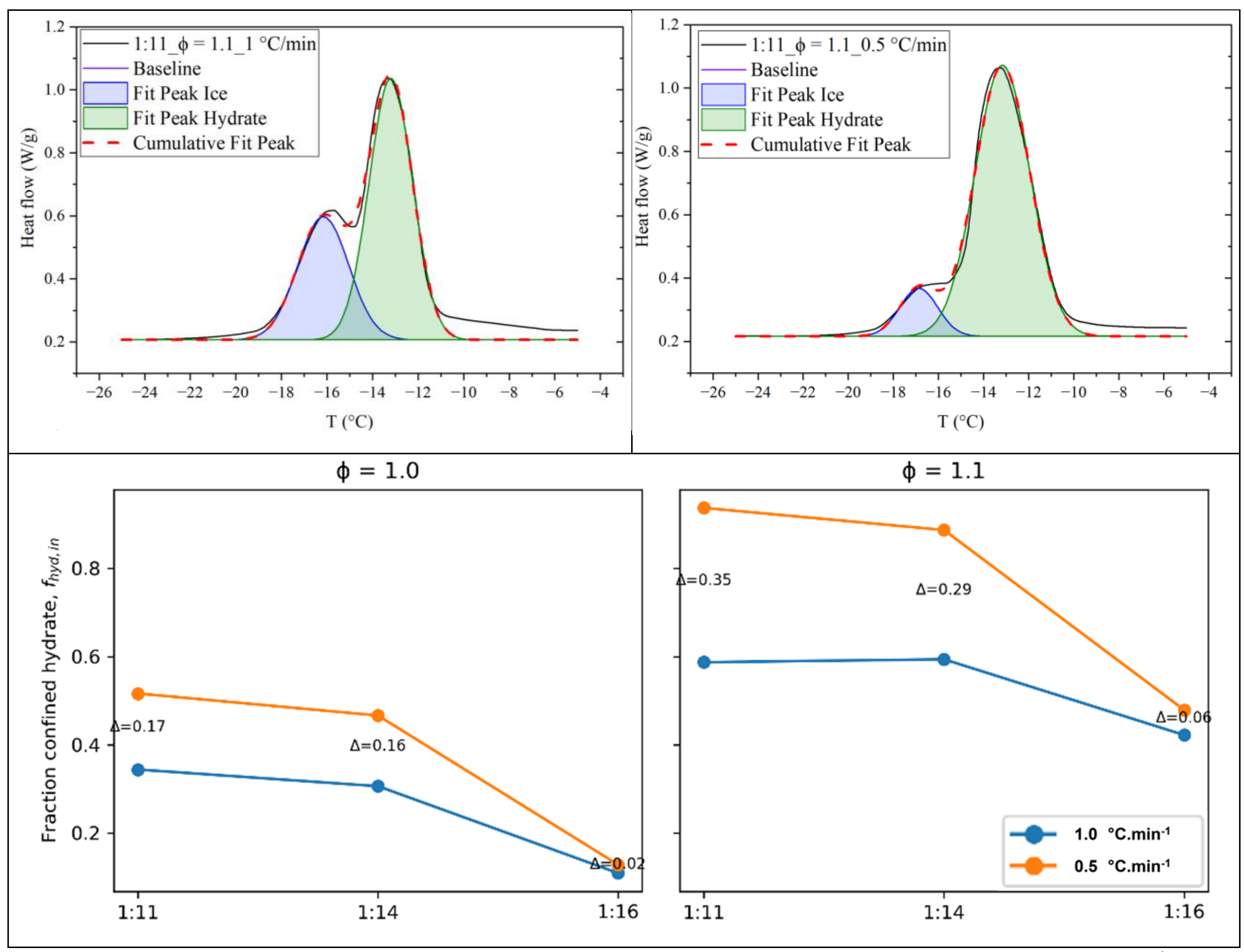

*Fig. 5. Top: Example peak decomposition (1:11, φ ≈ 1.1) on reheating at 2 °C.min$^{-1}$ after precooling at 1.0 (left) and 0.5 °C.min$^{-1}$ (right); the total endotherm is resolved into confined-ice and confined-hydrate contributions with $R^2 \geq 0.99$ of fitting quality. Bottom: Ratio of the confined hydrate enthalpy to the total endothermic enthalpy ($f_{hyd} = \frac{\Delta H_{hyd}}{\Delta H_{hyd}+\Delta H_{ice}}$) for φ=1.0 and φ=1.1 as a function of composition (1:11, 1:14, 1:16), comparing the two cooling rates.*

### 3.4. Effects of thermal cycling on the clathrate formation

We found that the presence of an excess phase outside the porous medium significantly affects the types of phases formed within the porosity, and particularly, the ratio of confined clathrate to confined ice. This raises the question of whether this ratio is fully determined, and thus reproducible, once the composition of the bulk solid phases and the confined solution is fixed. To address this question, we conducted thermal cycling experiments in which the confined phases were melted and then reformed, while the external bulk solid phases remained intact.

Practically, we selected three compositions (THF:$H_2O$ = 1:11, 1:14, 1:16) at the highest filling fraction ($\varphi$ = 1.1) and modified the DSC program so that each sample experienced a controlled cool → heat → re-cool → re-heat sequence. The first cooling was at 1 °C/min to the low-temperature limit (-40 °C), followed by a short isotherm, then a partial reheating at +2 °C/min to a temperature just above the confined-ice melt and the confined-hydrate dissociation (-8 °C). The sample was then cooled again after 1 min at −1 °C/min and, after a 1 min hold, reheated at +2 °C/min to complete the cycle (green = first cool/heat, red = re-cool, blue = re-heat in Fig. 6).

In all three mixtures, the blue reheating traces reproduce the first-heat dissociation and melting temperatures within our instrumental resolution: the confined-hydrate endotherm remains near ~ −13 °C, the confined-ice melt stays at its Gibbs–Thomson depressed value, and, when excess liquid is present at $\varphi$ = 1.1, the external/bulk-like hydrate near ~ +4 °C is likewise invariant with respect to previous experiments. Whereas the exotherms on both the first cooling (green) and the re-cooling (red) show the usual run-to-run scatter characteristic of heterogeneous formation inside the mesopores. As shown in Fig. 6, there is no systematic shift in the subsequent melting/dissociation temperatures, nor any measurable drift in peak shapes beyond minor enthalpy variations expected from stochastic formation, confirming that peak positions are state functions fixed by curvature, not by prior kinetics.

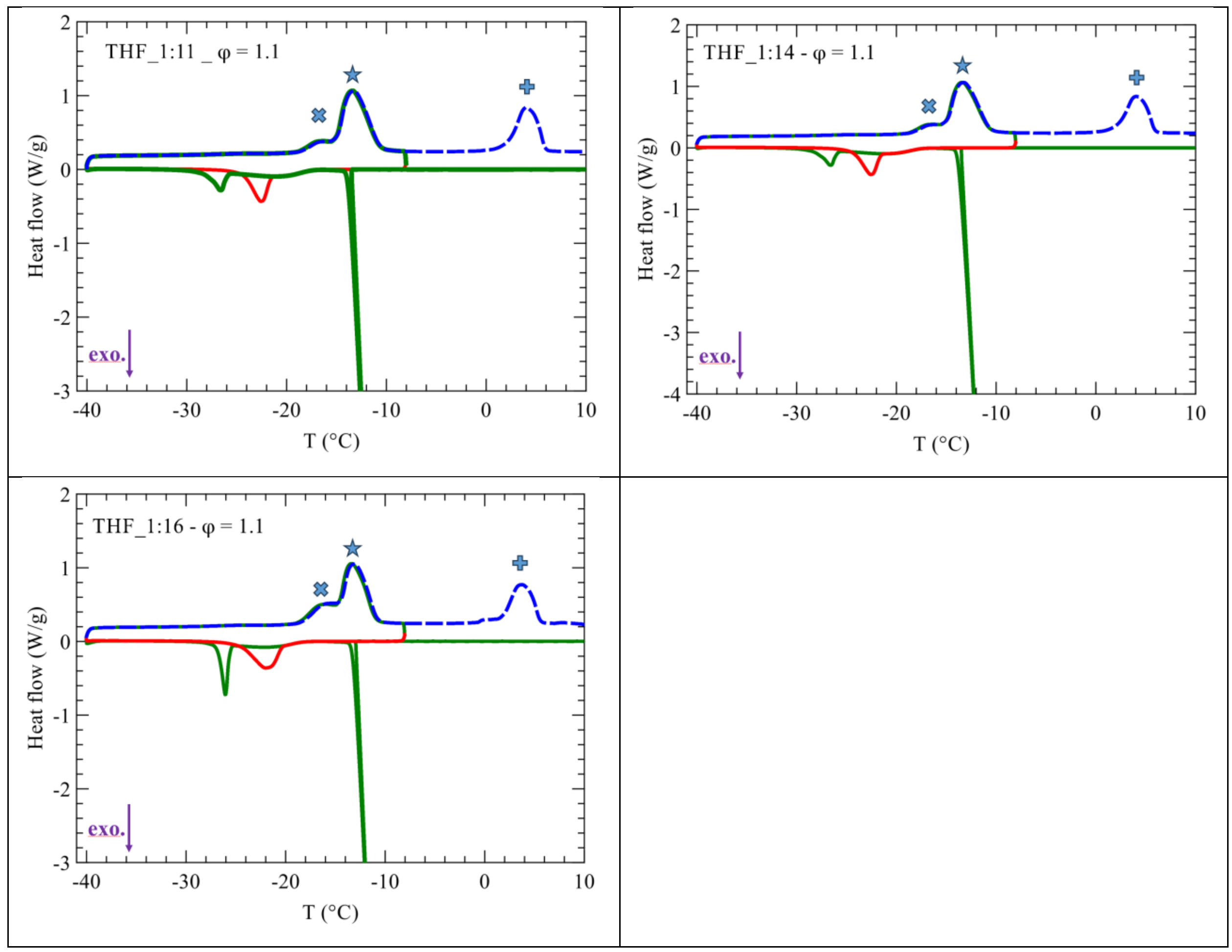

*Fig. 6. Reproducibility test for confined THF-water at φ ≈ 1.1 and three compositions (1:11, 1:14, 1:16). DSC sequence: cool to −40 °C at −1 °C.min$^{-1}$, 1 min hold; reheat to −8 °C at +2 °C.min$^{-1}$; re-cool at −1 °C.min$^{-1}$ (1 min hold), and final re-heat at +2 °C.min$^{-1}$ (green = first cool/heat, red = re-cool, blue = re-heat). Reheating temperatures are invariant: confined-ice melt (GT-depressed) and confined-hydrate dissociation (~ −13 °C), and bulk-like hydrate (~+4 °C) match the first-heat values. Cooling exotherms of confined solution vary stochastically, affecting peak areas but not melting/dissociation temperatures. On heating, the symbols show the different phase transitions: × relates to the melting of confined ice, ★ to the dissociation of confined hydrate, and + to the dissociation of the bulk/external hydrate phase.*

The invariance of the melting and dissociation temperatures across cycles is exactly what is expected when the relevant intensive variables are unchanged between scans. Once the sample is heated just past the confined-ice melt and the hydrate dissociation, the confined liquid re-equilibrates inside the pores while the outside solid is in crystalline form, and the composition outside the pore is constant. Upon the next cycle, the confined solution encounters the same capillarity-set equilibria. As shown in equation (2), those temperatures are set only by the contribution of a curvature (capillarity) term that scales with $\frac{2\gamma_{sl}\nu_s}{r\Delta H_m}$ (i.e., classical Gibbs-Thomson framework for confined solutions).[38, 40, 45] Because the pore radius ($r$) remains unchanged during these temperature changes, the confinement conditions are the same in each cycle. Therefore, the temperatures of the endothermic peaks remain unchanged upon reheating, while only their areas vary slightly depending on how much hydrate and ice formed during the preceding cooling step.

$$T_{bulk}^{0} - T_{r}^{w} = \Delta T_{m}(r) = \frac{2\,\gamma_{sl}\,\nu_{s}\,T_{bulk}^{0}}{\Delta H_{m}\,(r-t)} \qquad \textit{(equation 2)}$$

Where $T_{r}^{w}$ is the liquidus temperature in pores of radius ($r$), $T_{bulk}^{0}$ corresponds to the bulk liquidus temperature of the corresponding phase, $\gamma_{sl}$ shows solid-liquid interfacial free energy, $\nu_{s}$ is the molar volume of liquid or ice; the difference is negligible here, $\Delta H_{m}$ indicates molar enthalpy of fusion/dissociation, and t is a thin non-freezing interfacial layer at the silica wall (effective radius $r_{eff} = r - t$).

We next probed whether the melting and dissociation temperatures depend on the thermal itinerary when we interrupt the cycle after confined-ice melting but before hydrate dissociation, and whether changing the re-cooling rate alters the subsequent thermogram profiles. From a practical point of view, we only chose 1:14 mol:mol at φ ≈ 1.1, then ran a three-pass program: −0.5 °C.min$^{-1}$ → +2 °C.min$^{-1}$ → −0.5 °C.min$^{-1}$ → +2 °C.min$^{-1}$ → −1.0 °C.min$^{-1}$ → +2 °C.min$^{-1}$. The idea for doing such measurements came from the fact that the sensitivity of THF's hydrogen bonding to THF guest molecules in sII small cages may suggest that local composition can tune stability and dynamics.[55] Complementary Raman studies at low temperature show that the THF guest experiences cage-imposed restrictions and symmetry changes, reinforcing the link between microscopic interactions and macroscopic thermophysical response.[56]

In Fig. 7, the black dashed line is the first, complete cool/heat; the green one is the second cool, followed by a partial re-heat that stops just above the confined-ice melt (one-minute isotherm); the red line is the subsequent re-cool, and the blue dashed line is the final full re-heat. Two robust features emerge. First, the reheated curves (blue) reproduce, within our experimental uncertainty, the same temperatures for the Gibbs–Thomson depressed confined-ice melt and for the confined-hydrate dissociation as observed on the first heat (black). Second, switching the recooling rate from −0.5 to −1.0 °C.min$^{-1}$ when the excess composition is fixed because of the crystallization, changes neither the *amount* of phase formed (peak areas via stochastic nucleation) nor shifts melting/dissociation temperatures, which shows again that the liquidus in a pore is fixed by capillarity (curvature *r*). Our protocol heats only to just above the confined-ice melt, allowing the in-pore liquid to re-equilibrate compositionally while leaving the solid hydrate fraction present. Because the composition of the confined liquid is re-established under the same confinement conditions after each cycle, the subsequent reheating follows the same liquidus and hydrate–liquid coexistence conditions; accordingly, the peak temperatures remain unchanged.

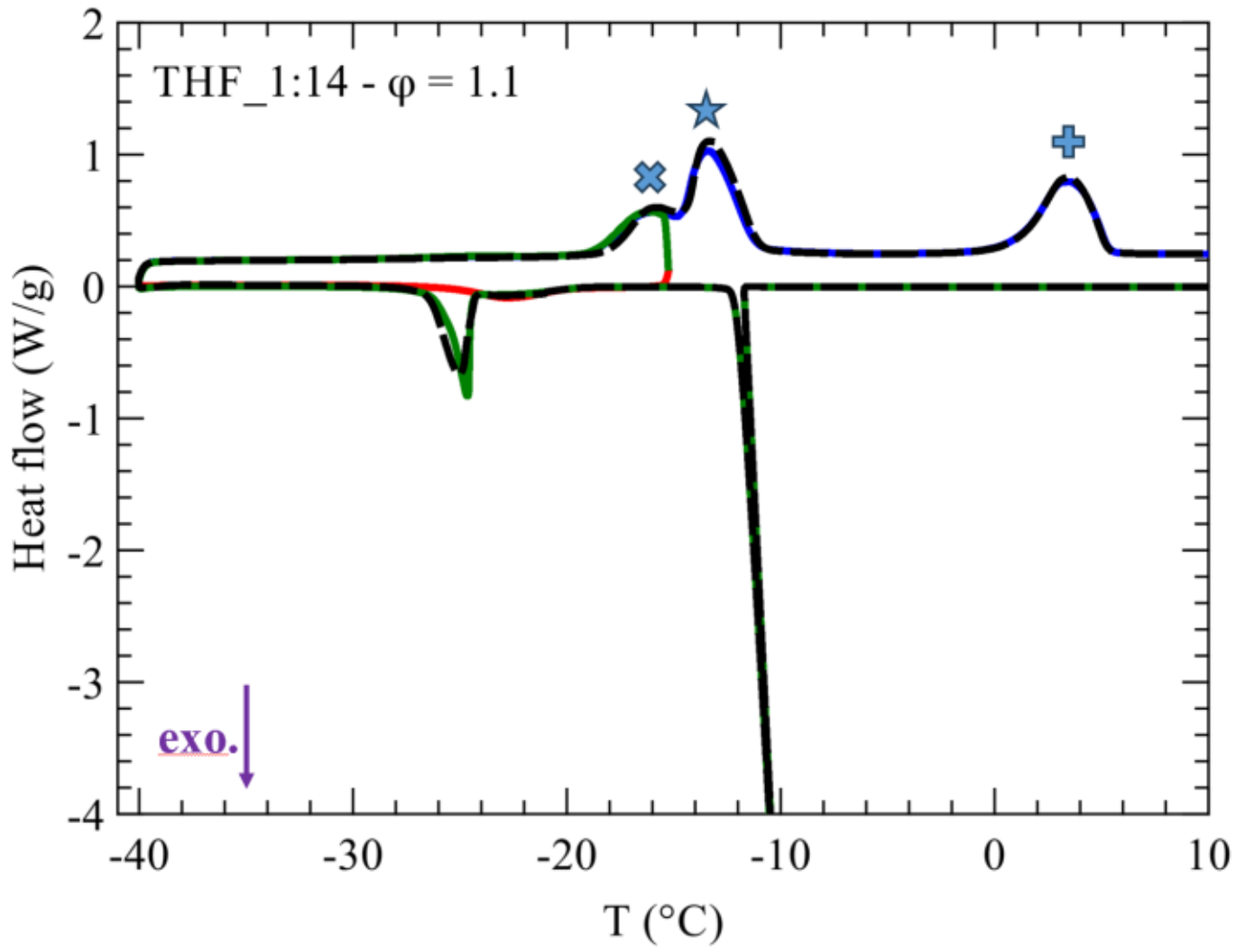

*Fig. 7. Thermal itinerary test at φ = 1.1 for THF:$H_2O$ = 1:14. Program: −0.5 → +2 → −0.5 → +2 → −1.0 → +2 °C.min$^{-1}$ (cool/heat/cool/heat/cool/heat). Traces: black dashed = first full cool/heat; green = second cool + partial re-heat to just above the confined-ice melt (1-min hold); red = re-cool; blue = final full re-heat. Reheating temperatures are invariant. The confined-ice melt (GT depressed) and confined-hydrate dissociation match the first-heat values expected from the Gibbs-Thomson relation at fixed curvature.*

## 3.5. Classical Gibbs–Thomson effect on confined ice and THF hydrate transitions

To quantify the capillarity control that underpins the DSC results, we measured the depression of the equilibrium temperatures for confined phases at φ ≈ 1.1 and THF:$H_2O$ = 1:14. For each silica mesoporous materials batch with distinct nominal pore radius r, we determined $\Delta T_m = T_{bulk} - T_r$ for (i) the Gibbs–Thomson depressed melting of confined ice and (ii) the dissociation of the confined THF sII hydrate. Plotting $\Delta T_m$ against $r^{-1}$ yields clean linear trends for both transitions (see Fig. 8), with coefficients of determination $R^2 \approx 0.99$. This is the classical Gibbs–Thomson response to curvature; the slope $A = \frac{2\,\gamma_{sl}\,\nu_s\,T^0_{bulk}}{\Delta H_m}$ encodes the interfacial-free-energy/latent-heat balance, and the larger slope for hydrate reflects its greater $^{v_m}/_{\Delta H_m}$ (with a modest $\gamma_{sl}$ difference) relative to ice.

Fitting $\Delta T_m = {}^{A}/_{r} + B$ gives $A_{hydrate} \approx 81.5 \pm 2.9$ K.nm and $A_{ice} \approx 73.1 \pm 4.2$ K.nm for our samples (nominal manufacturer radii). Then, we tried to give a consistent picture of the slope variation by the variation of the lever, which is the $^{v_m}/_{\Delta H_m}$. For ice (Ih, 273.15 K), $\Delta H_m = 6.01$ KJ/mol and $v_m = 19.65$ cm$^3$/mol give $^{v_m}/_{\Delta H_m} = 3.27$ cm$^3$/KJ. For sII THF.17$H_2O$ at 277 K, $\Delta H_m = 99.5$ KJ/mol and $\rho = 0.95$ g/cm$^3$ give $v_m = 398.3$ cm$^3$/mol and $^{v_m}/_{\Delta H_m} = 4$ cm$^3$/KJ.

If $\gamma_{sl}$ were identical, the expected slope ratio would be $\frac{A_{hyd}}{A_{ice}} \approx \frac{T_0^{hyd}}{T_0^{ice}} \frac{(v_m/\Delta H_m)_{hyd}}{(v_m/\Delta H_m)_{ice}} \approx 1.22$. However, experimentally (Fig. 8), we obtain $\frac{A_{hyd}}{A_{ice}} = 1.11$. Reconciling the measured ratio with the thermodynamic scaling yields $\frac{\gamma_{sl}^{hyd}}{\gamma_{sl}^{ice}} \approx \frac{1.11}{1.22} \approx 0.91 \pm 0.05$, indicating that the hydrate solid–liquid interfacial free energy is about 10 % lower than that of ice in the same matrix. The hydrate nevertheless shows the larger GT slope ($A_{hyd} > A_{ice}$) because its thermodynamic lever $^{v_m}/_{\Delta H_m}$ is larger; the smaller $\gamma_{sl}$ simply reduces the slope ratio from 1.22 to the measured 1.11.

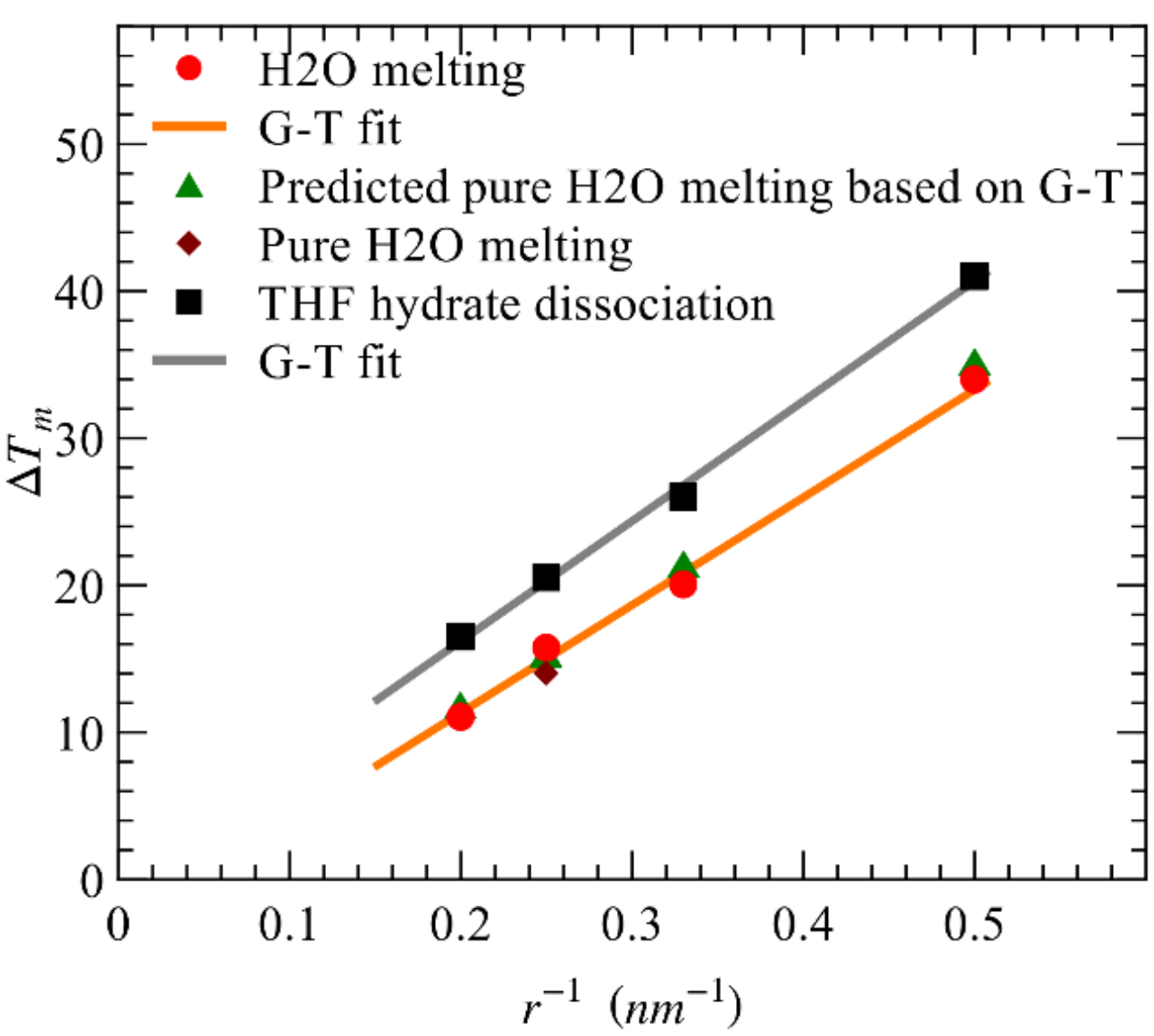

*Fig. 8. Gibbs-Thomson scaling for confined THF-water (1:14, φ ≈ 1.1). Temperature depression, $\Delta T_m$, vs inverse pore radius $r^{-1}$: red circles = confined $H_2O$ melting in the THF-water mixture; black squares = confined sII THF–hydrate dissociation. Solid lines are GT fits (orange = $H_2O$,*

*gray = hydrate; ($\Delta T_m = 1/r$). For reference, green triangles show the GT-predicted $\Delta T_m$ for pure water using the same pore sizes,[38] and brown diamonds are measured pure-water melting ($\varphi \approx 1.1$) in SBA-15 shown in Fig. S3a.*

A fully quantitative, classical GT approach for a pure substance explains the low-temperature endotherm in both pure water and THF-water runs. For ice melting in cylindrical pores, corrected for a thin nonfreezing interfacial layer, $\Delta T_m = \frac{C_{GT}}{r-t}$, with $C_{GT} = \frac{2\,\gamma_{sl}\,\nu_s\,T^0_{bulk}}{\Delta H_m}$ = 51.9 K.nm and t = 0.6 nm for hydrophilic silica.[38] As shown earlier, the low-temperature endotherm assigned to "confined ice" reports the equilibrium ice-water coexistence under curvature, because it is assumed that all THF is in sII hydrate, so it is immobilized and does not participate in the melt. Accordingly, the observed depression $T_{bulk} - T_{confined}$ matches perfectly the pure-water GT prediction for the same pore size. Using the measured $\Delta T_m = 14.65$ K for the pure water in SBA-15 (see Fig. S3a), indicating by the brown diamond in Fig. 8, we infer $r_{eff} = {}^{C_{GT}}/_{\Delta T_m} = 3.52\ nm$ and $r \approx r_{eff} + t = 4.1$ nm (D $\approx$ 8.2), consistent with $N_2$-sorption. For the four pore sizes studied here, the predicted depressions follow $\Delta T_m(r_i) = \frac{51.9}{r_i - 0.6}$ ($r_i$ in nm), providing a parameter-free benchmark that quantitatively rationalizes the aligned ΔT values we observe for both pure water and THF-water at the confined-ice transition. In practice, this means we can read the low-temperature peak as a curvature-set equilibrium marker with $\Delta T_m$ predicted from geometry alone, place the predicted values beside the measured depressions for each batch to show one-to-one agreement, and then treat the hydrate endotherms separately as reflecting phase amounts shaped by the thermal itinerary rather than shifting the ice liquidus itself.

For comparison, a melting depression of about 8.5 K was reported for ice and THF hydrate confined in Vycor with a pore radius of 4.2 nm.[37] In our case, the Gibbs–Thomson analysis predicts a substantially larger depression, about 14 K, for cylindrical pores of comparable radius. This larger value is consistent with established studies of water confined in ordered mesoporous silicas such as MCM-41 and SBA-15, which also report strong curvature-induced melting depressions.[39] The slower heating rate used in the Vycor study (0.2 K/min) is unlikely to account, by itself, for a difference of nearly 6 K in the apparent melting temperature. A more plausible origin of the discrepancy is the different nature of the porous network, in particular the broader pore-size distribution and interpore connectivity of Vycor. In addition, the cited paper shows only part of the melting endotherm, which makes it difficult to determine exactly how the melting temperature was evaluated and therefore prevents a strict comparison with the present DSC analysis.

Importantly, we do not extrapolate this GT relation to smaller pores than those accessed experimentally. As r approaches the nonfreezing-layer thickness t, the crystalline core $r_{ice} = r - t$ shrinks to only a few lattice spacings and the classical GT assumptions (a bulk-like solid phase with a well-defined curved interface and size-independent thermodynamic parameters) become questionable; in this regime, crystallization itself can be strongly hindered or suppressed.[57] This perspective is consistent with experimental and theoretical evidence for a lower size limit of bulk-like ice I order ("the end of ice I") in nanoscopic water systems, where crystalline signatures disappear below a characteristic cluster size scale.[57]

At the same time, the GT geometry specifies the size of the crystalline ice core at melting, $r_{ice} = r - t$, with the frozen cross-sectional fraction $(\frac{r_{ice}}{r})^2 = (1 - \frac{t}{r})^2$: 3.40 nm (6.80 nm; 72.3%) for r = 4.00 nm. This limits the lateral correlation length of any crystalline phase formed in the mesochannels. Given that sII hydrate has a cubic unit cell with a lattice parameter of about 1.7 nm,[8, 14] only a few unit cells can fit across the pore diameter in the mesopores studied here. As a result, crystalline coherence in the direction perpendicular to the pore axis is necessarily limited by confinement and by the presence of disordered interfacial layers near the silica surface. This geometrical constraint helps explain why hydrate signatures can still be detected in SBA-15 for pore radii down to about 2 nm, whereas for smaller pores the development of bulk-like crystalline order becomes increasingly unfavorable.

Thermodynamically, this same curvature $\propto 1/(r - t)$ that depresses the melting/dissociation temperature also enhances surface contributions to the free energy; structurally, it should manifest as broadened Bragg features without systematic peak-position shifts relative to bulk sII. We now test these predictions by wide-angle X-ray scattering (WAXS), comparing bulk THF-water hydrate with the product formed inside SBA-15.

### 3.6. Wide-angle X-ray scattering (WAXS) of bulk and confined THF-water hydrate

To identify the crystalline phase detected calorimetrically and test whether confinement perturbs its lattice, we conducted WAXS measurements of THF-water in bulk solution and in SBA-15 (d =

10 nm; r = 5 nm) over −30 to +10 °C. Diffractograms for the SBA-15 sample at excess filling ($\varphi > 1.0$) and for the bulk reference were measured over the same *Q*-range and reduced identically. Figure 9b overlays representative patterns from the SBA-15 sample with calculated line positions for the cubic clathrate structure II (sII) and with the broad amorphous SBA-15 background.

As shown, across the full temperature range, the Bragg peaks in both environments coincide one-for-one with sII reflections. All intense lines, most prominently those between $Q \approx 1.4$ and 2.2 $Å^{-1}$, appear at the expected positions, with no additional peaks attributable to ice Ih or alternative hydrate structures within the detection limit. The overlay demonstrates that the THF hydrate shares the same d-spacings as the sII reference within instrumental resolution. Thus, the guest-stabilized water framework is unchanged in SBA-15 relative to the bulk. In the patterns of SBA-15 sample, the clathrate peaks sit on top of a broad silica contribution centered near $Q \approx 1.6$–2.2 $Å^{-1}$, subtracting this background leaves the sII reflections at unchanged positions.

It is important to note, however, that the diffraction pattern measured for the SBA-15 sample at excess filling is a superposition of contributions from hydrate formed inside the pores and hydrate formed outside the pores, together with the broad amorphous background of the silica matrix. Therefore, unlike DSC, WAXS does not allow the in-pore and out-pore hydrate populations to be separated directly. Its role here is instead to determine whether any additional crystalline phase or detectable structural change appears in the SBA-15 sample relative to the bulk reference.

Consistent with the calorimetric signature, the confined system exhibits two dissociation intervals: −30 to −5 °C (in-pore) and 2 to 10 °C (out-pore) hydrate. In fact, temperature modulates peak intensities and widths but not positions. Within experimental uncertainty, sII peak positions are invariant from −30 to +10 °C in both bulk and confinement (in-pore and out-pore hydrate), indicating that any thermal expansion or composition-driven lattice change is below our resolution. Under confinement, the reflections are systematically broader, while remaining coincident in *Q*, a sign of finite coherent domain size and/or microstrain imparted by the mesopore walls rather than a change in structure or lattice parameter.

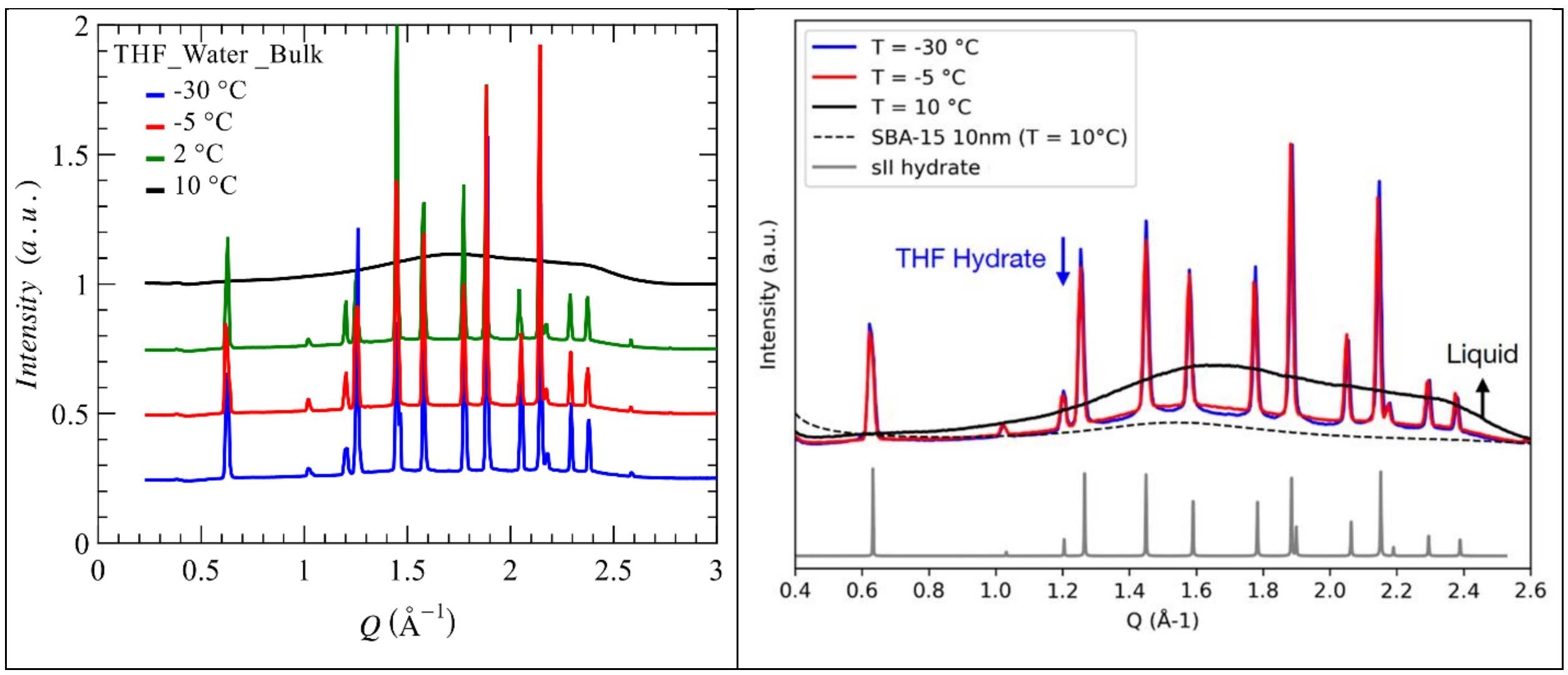

*Fig. 9. Wide-angle X-ray scattering of THF-water hydrate in bulk (left) and confined in SBA-15 (φ > 1.0) (right) from −30 to +10 °C. All Bragg peaks index to cubic sII clathrate; no ice Ih reflections are detected in the shown window. In the right panel, the overlay of confined patterns with the SBA-15 background (dashed) and the sII reference (sticks), confirming identical d-spacings for confined and bulk hydrates.*

In short, WAXS shows that the confined product is crystallographically the same cubic sII clathrate as in bulk: peak positions match within resolution and only moderate line broadening is observed on the amorphous silica background. Accordingly, the lower dissociation/melting temperatures seen by DSC reflect capillarity in the mesopores, not a change of crystal structure (i.e., confinement leaves the structure and lattice spacings intact).

## Conclusion

In this study, we have used differential scanning calorimetry and wide-angle X-ray scattering (WAXS) to establish a consistent thermodynamic picture for clathrate hydrate formation of Tetrahydrofuran (THF)-water in bulk and mesoporous SBA-15, in which equilibrium and kinetics play distinct, nonoverlapping roles. On heating, the Gibbs–Thomson depressed confined-ice melt (= −14.7 °C) and the in-pore sII hydrate dissociation (= −13.2 °C) recur almost at the same temperatures regardless of itinerary, demonstrating that the liquidus in a cylindrical pore is fixed solely by capillarity via the Gibbs-Thomson relation, rather than any kinetic history. A key finding of this study is that confinement preferentially promotes hydrate formation over water crystallization. In the THF–water system, this effect is reinforced by the antifreeze role of THF, which further depresses the crystallization temperature of water and thereby facilitates hydrate formation under milder conditions.

Hydrate appears inside the pores only when the network is effectively percolated (φ = 1.0 – 1.1 $cm^3/g$) and the mixture is sufficiently THF-rich (≥ 1:16); the nominal 1:17 series lacks a confined-hydrate signal because small, reproducible THF losses during loading shift the effective composition water-rich. Within the admissible window set by composition and connectivity, the cooling itinerary governs amounts, not temperatures. By changing the precooling rate from 1.0 to 0.5 °C/min, on heating, the invariant positions of the hydrate dissociation and ice-melting peaks report equilibrium constraints (chemical potentials) fixed by composition and, under confinement, by capillarity (Gibbs-Thomson effect); on cooling, rate-dependent exotherms track stochastic nucleation and growth without shifting equilibrium temperatures, an interpretive stance consistent with the stochastic picture and with the thermodynamic phase boundaries. However, a slower cooling rate increases the confined-hydrate fraction $f_{hyd} = \frac{\Delta H_{hyd}}{\Delta H_{hyd}+\Delta H_{ice}}$ by about 0.29 – 0.35 at φ=1.1 for the 1:11 and 1:14 compositions. In the same samples, the hydrate enthalpy increases by approximately 54–60%, whereas the ice contribution decreases by about 70–85%. These changes show that the cooling protocol mainly affects the relative amounts of hydrate and ice formed during cooling, while the equilibrium transition temperatures remain unchanged. Wide-angle X-ray scattering further shows that the confined hydrate is crystallographically identical to bulk sII hydrate. Under confinement, the Bragg peaks are slightly broadened, but their positions remain unchanged. This indicates that confinement affects the finite size and structural coherence of the crystallites, without modifying the hydrate structure itself. Therefore, the calorimetric shifts observed under confinement are attributed to capillarity effects rather than to a change in clathrate structure or lattice spacing. Finally, the depressions of the confined-ice melting temperature and the confined-hydrate dissociation temperature both vary linearly with the inverse pore radius, as expected from the Gibbs–Thomson relation. This provides a direct quantitative link between pore size and calorimetric signatures, showing that the transition temperatures measured by DSC can be used to relate the observed thermal behavior to the characteristic size of the confinement.

All in all, this work disentangles thermodynamic effects, reflected in invariant peak temperatures, from kinetic effects, reflected in rate-dependent peak areas, and establishes a quantitative Gibbs–Thomson framework linking DSC transition temperatures to pore size in nanoconfined hydrate systems. These results should guide the design and interpretation of hydrate formation in nanoconfinement, including promoter-assisted gas-storage or separation schemes where phase selection must be steered without altering crystal structure. As a perspective for future research, it would be insightful to apply the systematic approach used in the present work to the case of radically different hydrate-forming molecules. Among these, cyclopentane stands out as particularly intriguing due to its complete immiscibility with liquid water, which results in the formation of interfacial hydrates with a fibrous morphology that initiate from the surface of porous media.[58, 59]

## Supplementary Material

See the Supplementary Material for additional experimental details and supporting data, including a schematic of the SBA-15 synthesis, structural characterization of the SBA-15 series by SAXS and $N_2$ adsorption–desorption isotherms, DSC pan mass variations before and after measurements, determination of the SBA-15 pore volume from DSC measurements of pure water, and additional analysis of the cooling-rate dependence of the confined-hydrate enthalpy fraction.

## Acknowledgments

This research was carried out as part of the ANR FIDELIO project (ANR-22-CE50-0002), for which we express our gratitude. The authors gratefully acknowledge Daniel Dudzinski for his technical assistance and for maintaining the X-ray diffraction laboratory infrastructure that supported this work.